\documentclass[conference,letterpaper]{IEEEtran}

\usepackage[utf8]{inputenc} 
\usepackage[T1]{fontenc}
\usepackage{url}
\usepackage{ifthen}
\usepackage{cite}
\usepackage[cmex10]{amsmath} 

\usepackage{graphics, epstopdf, graphicx, epsfig, psfrag, epsf, subfigure}
\usepackage{amssymb, amsfonts, amsma th, times}
\usepackage{relsize}
\usepackage{multicol,balance, verbatim}   
\usepackage{textcomp, mathrsfs, stmaryrd, setspace}
\usepackage{amssymb}
\usepackage{wrapfig}
\usepackage{xcolor}
\usepackage{cite}
\usepackage{xspace}

\newcommand{\ie}{{\em i.e., }}

\newcommand{\eg}{{\em e.g., }}

\newtheorem{theorem}{Theorem}

\newtheorem{corollary}[theorem]{Corollary}
\newtheorem{definition}{Definition}
\newtheorem{example}{Example}

\newcommand{\Nset}{\mathcal{N}}

\newcommand{\ACMM}{{\tt {ACM$^2$}}\xspace}

\begin{document}
\title{Adaptive Coding for Matrix Multiplication at Edge Networks} 


\author{%
  \IEEEauthorblockN{Elahe Vedadi and Hulya Seferoglu}
  \IEEEauthorblockA{University of Illinois at Chicago\\
                    Email: \{evedad2, hulya\}@uic.edu}
}


\maketitle

\begin{abstract}
Edge computing is emerging as a new paradigm to allow  processing data at the  edge  of  the  network, where data is  typically  generated  and  collected,  by exploiting  multiple devices at the edge collectively. However, exploiting the potential of edge computing is challenging mainly due to the heterogeneous and time-varying nature of edge devices. Coded computation, which advocates mixing data in sub-tasks by employing erasure codes and offloading these sub-tasks to other devices for computation, is recently gaining interest, thanks to its higher reliability, smaller delay, and lower communication cost. In this paper, our focus is on characterizing the cost-benefit trade-offs of coded computation for practical edge computing systems, and develop an adaptive coded computation framework. In particular, we focus on matrix multiplication as a computationally intensive task, and develop an adaptive coding for matrix multiplication (\ACMM) algorithm by taking into account the heterogeneous and time varying nature of edge devices. \ACMM dynamically selects the best coding policy by taking into account the computing time, storage requirements as well as successful decoding probability. We show that \ACMM improves the task completion delay significantly as compared to existing coded matrix multiplication algorithms.


\end{abstract}


\section{Introduction}
Massive amount of data is generated at edge networks with the emerging Internet of Things (IoT) including self-driving cars, drones, health monitoring devices, etc. 
Transmitting such massive data to the centralized cloud, and expecting timely processing are not realistic with limited bandwidth between an edge network and centralized cloud. 
We consider a distributed computing system, where computationally intensive aspects are distributively  processed at the end devices with possible help from edge servers (fog) and cloud.  However, exploiting the potential of edge computing is challenging mainly due to the heterogeneous and time-varying nature of edge devices.

Coded computation is an emerging field, which studies the design of erasure and error-correcting codes to improve the performance of distributed  computing through ``smart'' data redundancy. This breakthrough idea has spawned a significant effort, mainly in the information and coding theory communities \cite{SpeedUp-journal, Tradeoff-journal}. According to distributed computation, a \emph{master} device divides computationally intensive aspects/tasks into multiple smaller sub-tasks, and offloads each of them to other devices (end devices, edge servers, and cloud), called \emph{workers}, for computation. Coded computation (\eg by employing erasure codes such as Reed Solomon codes \cite{macwilliams1977theory,lin1983error}), on the other hand, encodes the data in the sub-tasks, and offloads these coded sub-tasks for computation. The next example demonstrates the potential of coded computation for matrix multiplication.

 \begin{example} \label{ex1}
 Consider a setup where a master  wishes to offload a matrix multiplication $C = A ^{T} B$ task to three workers. Assume $A$ and $B$ are $K \times K$ matrices and matrix $A$ is divided into two sub matrices $A_1$ and $A_2$, which are then encoded using a $(3,2)$ Maximum Distance Separable (MDS) code, which is further explained in Section~\ref{sec:system}, 
 to give  $Z_1=A_1$, $Z_2=A_2$ and $Z_3=A_1+A_2$, and sends each to a different worker. When the master receives the computed values (\ie $Z_i^TB$) from at least two out of three workers, it can decode its desired task, which is the computation of $A^T B$. The power of coded computations is that it makes $Z_3=A_1+A_2$ acts as an extra task that can replace any of the other two tasks if they end up straggling or failing. \hfill $\Box$
  \end{example}
  
Significant effort is being put on constructing codes for fast and distributed matrix-vector multiplication \cite{SpeedUp-journal, ferdinand2016anytime}, matrix-matrix multiplication \cite{yu2017polynomial, lee2017high, fahim2017optimal, yu2018straggler}, dot product and convolution of two vectors \cite{dutta2016short, dutta2017coded}, gradient descent \cite{tandon2017gradient, halbawi2018improving, pmlr-v80-raviv18a}, distributed optimization \cite{karakus2017encoded, karakus2017straggler}, Fourier transform \cite{yu2017coded}, and linear transformations  \cite{yang2017computing}. The trade-off between latency of computation and load of communication for data shuffling in MapReduce framework is characterized in \cite{Tradeoff-journal}, and optimum resource allocation algorithm is developed in \cite{yu2017optimally}. This coding idea is extended for cellular networks \cite{li2017scalable}, multistage computation \cite{li2016coded}, and heterogeneous systems \cite{kiamari2017heterogeneous, reisizadeh2019coded}.

Our focus in this work is on matrix multiplication, where a master device divides its matrix multiplication computations into smaller tasks and assigns them to workers (possibly including itself) that can process these tasks in parallel. Product \cite{lee2017high}, polynomial  \cite{yu2017polynomial}, and  MatDot (and its extension PolyDot) codes \cite{fahim2017optimal} are recently developed for matrix multiplication. Their main focus is to minimize/optimize the recovery threshold, which is the minimum number of workers that the master needs to wait for in order to compute matrix multiplication ($C = A^T B$ in Example \ref{ex1}). Although this metric is good for large scale computing systems in data centers, it fails short in edge computing, where other resources including computing power, storage, energy, networking resources are limited. 

In this paper, we analyze the computing time, storage cost, and successful decoding probability of some existing codes for matrix multiplication. Then, we develop an adaptive coding for matrix multiplication (\ACMM) algorithm that selects the best coding strategy for each sub-matrix. 

We note that rateless codes considered in \cite{mallick2019rateless, keshtkarjahromi2018dynamic} also provide adaptive coded computation mechanisms against heterogeneous and time-varying resources. However, the coding overhead in rateless codes can be high in some scenarios, which makes adaptive selection of fixed-rate codes a better alternative. Multi-message communication by employing Lagrange coded computation is considered in \cite{ozfatura2020straggler} to reduce under-utilization due to discarding partial computations carried out by stragglers as well as over-computation due to inaccurate prediction of 
the straggling behavior. A hierarchical coded matrix multiplication is developed in \cite{kiani2019hierarchical} to utilize both slow and fast workers. As compared to \cite{ozfatura2020straggler, kiani2019hierarchical}, we propose an adaptive code selection policy for heterogeneous and time-varying resources. 
A code design mechanism under a heterogeneous setup is developed in \cite{kiamari2017heterogeneous, reisizadeh2019coded}, where matrix $A$ is divided, coded, and offloaded to worker by taking into account heterogeneity of resources. However, available resources at helpers  may vary over time, which is not taken into account in \cite{kiamari2017heterogeneous, reisizadeh2019coded}. Thus, it is crucial to design a coded computation mechanism, which is dynamic and adaptive to heterogeneous and time-varying resources, which is the goal of this paper.

We show that \ACMM improves the task completion delay significantly as compared to existing coded matrix multiplication algorithms. The following are the key contributions:
\begin{itemize}
    \item We provide computing time analysis of some existing codes designed for matrix multiplication including product \cite{lee2017high}, polynomial  \cite{yu2017polynomial}, and  MatDot codes \cite{fahim2017optimal}. 
    \item We characterize storage requirements of existing matrix multiplication codes \cite{lee2017high, yu2017polynomial, fahim2017optimal} at master and workers.
    \item We design \ACMM for an iterative procedure (\eg gradient descent) that selects the best code as well as the optimum number of partitions for that code at each iteration to minimize the average computing time subject to storage and successful decoding probability constraints.\footnote{We note that \ACMM is generic enough to work with any matrix multiplication codes although we consider a subset of existing codes in this paper such as  product, polynomial, and  MatDot codes.}  
    \item We evaluate the performance of \ACMM through simulations and show that \ACMM significantly improves the average computing time as compared to existing codes. 
\end{itemize} 


 
\section{\label{sec:system} Model, Background, and Motivation}

\subsection{Model}

\subsubsection{Setup}  We consider a master/worker setup at the edge of the network, where the master device offloads its computationally intensive tasks (matrix multiplication computations) to workers $w_n$, $n \in \Nset$ \big(where $\Nset \triangleq \{1,\dots,N\}$  \big) via device-to-device (D2D) links such as Wi-Fi Direct and/or Bluetooth. The master device divides a task (matrix) into smaller sub-matrices, and offloads them to parallel processing workers.

\subsubsection{Task Model} The master  wants to  compute  {\em functions} of its collected data, which is determined by the applications. We will focus on computing linear functions; specifically matrix multiplication $C=A^TB$, where $A \in \mathbb{R}^{L \times K}$, $B \in \mathbb{R}^{L \times K}$. Matrix multiplication forms an essential building block of many signal processing (convolution, compression, etc.) and machine learning algorithms (gradient descent, classification, etc.)   \cite{SpeedUp-journal}.
We consider an iterative procedure (gradient descent) where a matrix multiplication is calculated at each iteration. 

\subsubsection{Worker Model} The workers have the following properties:  \emph{(i) failures:}  workers may fail or ``sleep/die" or leave the network before  finishing their assigned computational tasks. \emph{(ii) stragglers:} workers will incur probabilistic delays in responding to the master. 

\subsubsection{Coding Model} We design and employ an adaptive coding for matrix multiplication (\ACMM) that selects the best coding strategy among repetition, MDS \cite{SpeedUp-conference,SpeedUp-journal}, polynomial \cite{yu2017polynomial}, MatDot \cite{fahim2017optimal}, and product codes \cite{lee2017high} by taking into account the  computing time, storage cost, and successful decoding probability of these codes. The master device divides the matrix $A$ into $p_{\pi}$ partitions (sub-matrices), where $\pi \in \{\text{rep}, \text{mds}\}$ for repetition and MDS codes. Both $A$ and $B$ matrices are divided into $p_{\pi}$ partitions, where $\pi \in \{\text{poly}, \text{matdot}, \text{pro} \}$ for polynomial, MatDot, and product codes.


\subsubsection{Delay Model}  Each sub-matrix transmitted from the master to a worker $w_n,\ n\in \Nset,$ experiences the following delays: (i) transmission delay for sending the sub-matrix from the master to the worker, (ii) computation delay for computing the multiplication of the sub-matrices, and (iii) transmission delay for sending the computed matrix from the worker $w_n$ back to the master. 
We model the composite delay using the shifted exponential distribution ($f(t)=\lambda e^{-\lambda(t-1)}$ for $t\geq 1$) \cite{liang2014tofec, SpeedUp-journal},  with $\lambda$ referred to as the straggling parameter and each sub-task with shifted-scaled exponential distribution ($f(t)=\alpha\lambda e^{-\alpha\lambda(t-\frac{1}{\alpha})}$ for $t\geq \frac{1}{\alpha} $) where the scale parameter, $\alpha$ is selected from $\alpha \in \{p_{\text{rep}}, p_{\text{mds}}, p^2_{\text{poly}}, p_{\text{matdot}}, p^2_{\text{pro}}\}$.

\subsection{\label{sec:existingCodes}Background on Existing Codes for Matrix Multiplication}

In this section, we provide a short background on existing coded computation mechanisms for matrix multiplication.



\subsubsection{Repetition Codes} The master device divides matrix $A$ column-wise into $p_{\text{rep}}$ parts, where $p_{\text{rep}}|N$, and generates $\frac{N}{p_{\text{rep}}}$ copies of each sub-matrix. Sub-matrix $A_i$, $i=1, \ldots, p_{\text{rep}}$ is transmitted to $\frac{N}{p_{\text{rep}}}$ workers as well as matrix $B$. Workers calculate $A_i^TB$, and return their calculation back to the master, which finishes matrix multiplication calculation when it receives $A_i^TB$, $\forall i=1, \ldots, p_{\text{rep}}$. 

\subsubsection{MDS Codes \cite{SpeedUp-conference,SpeedUp-journal}}
The master device divides the matrix $A$ column-wise to $p_{\text{mds}}$ partitions. An $(N, k_{\text{mds}})$ MDS code sets $k_{\text{mds}} = p_{\text{mds}}$ and codes  $k_{\text{mds}}$ sub-matrices into $N$ sub-matrices using existing MDS codes like Reed-Solomon codes. $A_i$, $\forall i=1, \ldots, N$ as well as $B$ are transmitted to $N$ workers. When the master device receives $k_{\text{mds}}$ $A_i^T B$ calculations, it can decode and calculate $A^TB$. 

\subsubsection{Polynomial Codes \cite{yu2017polynomial}} The master device divides $A$ and $B$ column-wise into $p_{\text{poly}}$ partitions, where $p_{\text{poly}}^2 \leq N$. The master constructs polynomials; $\alpha(n) = \sum_{j=1}^{p_{\text{poly}}} A_j^T n^{j-1}$ and $\beta(n) = \sum_{j=1}^{p_{\text{poly}}} B_j n^{(j-1)p_{\text{poly}}}$, and sends them to worker $W_n$, which calculates $\alpha(n) \beta(n)$ multiplication. When the master receives $k_{\text{poly}} = p_{\text{poly}}^2$ $\alpha(n) \beta(n)$ multiplication, decoding is completed.

\subsubsection{MatDot Codes \cite{fahim2017optimal}} Both matrices $A$ and $B$ are divided row-wise  into $p_{\text{matdot}}$ partitions, where $2p_{\text{matdot}}-1 \leq N$. The master constructs the polynomials; $\alpha(n) = \sum_{j=1}^{p_{\text{matdot}}}A_j^T n^{j-1}$ and $\beta(n) = \sum_{j=1}^{p_{\text{matdot}}} B_j n^{p_{\text{matdot}} - j}$, and sends them to worker $w_n$ for processing, where worker $w_n$ calculates $\alpha(n)\beta(n)$ multiplication. When the master receives $k_{\text{matdot}} = 2p_{\text{matdot}}-1$ results from workers, it can decode and calculate $A^TB$.

\subsubsection{Product Codes \cite{lee2017high}} Product codes extend MDS codes in a way that both $A$ and $B$ are partitioned column-wise and coded. In particular, both $A$ and $B$ are divided into $p_{\text{pro}}$ partitions, and these partitions are put into $p_{\text{pro}} \times p_{\text{pro}}$ array. Then, every row of the array is encoded with an $(\sqrt{N},p_{\text{pro}})$ MDS code, which results $p_{\text{pro}} \times \sqrt{N}$ array. This array is also coded with an $(\sqrt{N},p_{\text{pro}})$ MDS code, which results into $\sqrt{N}$-by-$\sqrt{N}$ array. Each coded sub-matrix in this array is sent to a worker (out of $N$ workers) for calculation.
Product codes are decodable if at least one entry of any possible sub-array with size larger than or equal to $(\sqrt{N}-p_{\text{pro}}+1) \times (\sqrt{N}-p_{\text{pro}}+1)$ is received successfully.



\begin{table*}[ht]
\vspace{3pt}
\centering
\caption{Comparison of product, polynomial, MatDot, MDS and repetition codes for matrix-matrix multiplication.}
\label{table:codes_comparison}
\begin{tabular}{|l|c|c|c|c|c|c|c|c|c|}
\hline
 & {\begin{tabular}[c]{@{}c@{}}Recovery \\ threshold \\ ($k$)\end{tabular}} & \begin{tabular}[c]{@{}c@{}}Computing \\ load \\ per worker ($\gamma$) \end{tabular} & \begin{tabular}[c]{@{}c@{}}Storage \\ load \\ per worker ($\mu$) \end{tabular} & \multicolumn{4}{c|}{\begin{tabular}[c]{@{}c@{}}Probability of \\ successful \\ computation ($\rho$) \end{tabular}}  \\ \cline{5-8} 
                  & & & &\multicolumn{1}{l|}{N=6} & \multicolumn{1}{l|}{N=7} & \multicolumn{1}{l|}{N=8} & \multicolumn{1}{l|}{N=9} \\ \hline
Product           & 6    & $\frac{K^3}{4}$     & $K^2 + \frac{K^2}{4}$    & N/A      & N/A    & N/A     & 0.63       \\ \hline
Polynomial        & 4    & $\frac{K^3}{4}$    & $K^2 + \frac{K^2}{4}$   & 0.67      & 0.82     & 0.91    & 0.96       \\ \hline
MatDot   & 3     & $\frac{K^3}{2}$      & $2K^2$    & 0.89         & 0.95          & 0.98         & 0.99       \\ \hline
MDS   & 2     & $\frac{K^3}{2}$      & $2K^2$    & 0.98         & 0.99          & 0.99         & 0.99       \\ \hline
Repetition   &  $\lfloor \frac{N}{2}+1 \rfloor$     & $\frac{K^3}{2}$      & $2K^2$    & 0.67         & N/A          & 0.73         & N/A       \\ \hline
\end{tabular}
\vspace{-5pt}
\end{table*}

\subsection{Motivation for Adaptive Coding} 
Assume a canonical setup, where $A$ and $B$ are $K \times K$ matrices and divided into two sub-matrices $A_0$, $A_1$, $B_0$, and $B_1$. The product codes divide matrices column-wise, \ie $A_i$ and $B_i$ are $K \times \frac{K}{2}$  
matrices, for $i \in \{0,1\}$, and use two-level MDS codes. Considering that $A_2 = A_0+A_1$ and $B_2 = B_0+B_1$, nine codes are constructed by $A_i^T B_i$, for $i,j \in \{0,1,2\}$. In polynomial codes the master device, by dividing matrices column-wise, creates polynomials $\alpha(n) = A_0 + A_1 n$ and $\beta(n) = B_0 + B_1 n^2$ for worker $w_n$, which multiplies $\alpha(n) \beta(n)$. MatDot follows a similar idea of polynomial codes with the following difference: $A$ and $B$ are divided row-wise. 

Table~\ref{table:codes_comparison} shows the recovery threshold ($k$) \cite{lee2017high, yu2017polynomial, fahim2017optimal}, computing load per worker ($\gamma$) (this is the simplified analysis presented in Section~\ref{sec:compTime} and further detailed in Appendix A), which shows the number of required multiplications, storage load per worker ($\mu$), which shows the average amount of memory needed to store matrices and their multiplication (as detailed in Section~\ref{sec:storage}), and probability of successful computation ($\rho$), which is calculated assuming that the failure probability of workers is $\frac{1}{3}$ and independent. 
As seen, although MDS is the best in terms of recovery threshold ($k$), it introduces more computing load per worker (because of partitioning only one of the matrices). Also, 
MatDot, MDS and repetition codes perform worse than polynomial and product codes in terms of storage load per worker. Product codes require at least $N=9$ workers due to their very  design, and for repetition codes the number of workers should be an even number, but MDS, polynomial and MatDot codes are more flexible. As seen, there is a trade-off among $\{k,\gamma, \mu, \rho \}$, which we aim to explore in this paper by taking into account the limited edge computing resources including computing power and storage. For example, if there is no constraint on the total number of workers, but only on computing load, we will likely select product or polynomial codes. Next, we will provide a computing time and storage analysis of existing codes, and develop \ACMM that selects the best code depending on edge constraints.

\section{\label{sec:acmm} Adaptive Coding for Matrix Multiplication} 
 \subsection{\label{sec:compTime}Computing Time Analysis}

Assuming a shifted-scaled exponential distribution as a computation delay model with $\lambda$ as the straggling parameter and $\alpha \in \{p_{\text{rep}}, p_{\text{mds}}, p^2_{\text{poly}}, p_{\text{matdot}}, p^2_{\text{pro}}\}$ as the scale parameter for each worker, average computing time for repetition codes $T_{\text{rep}}$ and MDS codes $T_{\text{mds}}$ are expressed \cite{SpeedUp-journal} as
\begin{align}
T_{\text{rep}} \approx \frac{1}{p_{\text{rep}}}\left(1+\frac{p_{\text{rep}}}{N\lambda}\log(p_{\text{rep}})\right),
\end{align}
\begin{align}
T_{\text{mds}} \approx \frac {1}{{p_{\text{mds}}}}\left(1+\frac{1}{\lambda}\log\left(\frac {N}{N-{k_{\text{mds}}}}\right)\right).    
\end{align}

\begin{corollary}
\label{polymatdottime}
The average computing time for polynomial codes $T_{\text{poly}}$ and MatDot codes $T_{\text{matdot}}$ is expressed as the following assuming that a shifted-scaled exponential distribution is used as a delay model.  
\begin{align}
T_{\text{poly}} \approx \frac{1}{{{p}^2_{\text{poly}}}}\left(1+\frac {1}{\lambda} \log\left(\frac {N}{N-{k_{\text{poly}}}}\right)\right),
\end{align}
\begin{align}
T_{\text{matdot}} \approx  \frac{1}{{p_{\text{matdot}}}}\left(1+\frac {1}{\lambda} \log\left(\frac {N}{N-{k_{\text{matdot}}}}\right)\right).  
\end{align}
\end{corollary}
{\em Proof:} 
The proof is provided in Appendix B. 
\hfill $\Box$


The product codes have different performance in two different regimes. In the first regime the number of  workers scales sublinearly with $p_{\text{pro}}^2$, \ie $N =  p_{\text{pro}}^2+\mathcal{O}(p_{\text{pro}})$, while in the second regime, the number of  workers scales linearly with $p_{\text{pro}}^2$, \ie $N = p_{\text{pro}}^2+\mathcal{O}(p_{\text{pro}}^2)$. The computing time analysis of product codes is provided in these two regimes next. 

\begin{corollary}
\label{producttime}
Assume a shifted-scaled exponential distribution as a delay model with $\lambda$ as the straggling parameter and $p^2_{\text{pro}}$ as the scale parameter for each worker, average computing time $T_{\text{pro}}$ for $(p_{\text{pro}}+\frac{\tau}{2}, p_{\text{pro}})^{2}$ product codes and $(p_{\text{pro}}+\frac{\tau}{2})^2$ workers, where $\tau$ is an even integer, as $p_{\text{pro}}$ grows to infinity, is expressed in the first regime as
\begin{align}\label{Tpro_corollary2}
    T_{\text{pro}} \approx \frac{1}{p_{\text{pro}}^2}\left(1+\frac{1}{\lambda}\log\left(\frac{p_{\text{pro}}+\frac{\tau}{2}}{c_{\tau/2+1}}\right)\right),
\end{align} where $c_{\tau/2+1} \approx (1+\tau/2)+\sqrt{(1+\tau/2)\log(1+\tau/2)}$ \cite{4313069}, \cite{PITTEL1996111}. Assuming the same delay distribution, the lower bound and upper bound of average computing time $T_{\text{pro}}$ for $(\sqrt{1+\delta} p_{\text{pro}}, p_{\text{pro}})^2$ product codes and $(1 + \delta)p^2_{\text{pro}}$ workers, for a fixed $\delta$, as $p_{\text{pro}}$ grows to infinity, is expressed in the second regime as
\begin{align}\label{Tpro_lower_corollary2}
    T_{\text{pro}}^{\text{low}} = \frac{1}{p_{\text{pro}}^2}\left(1+\frac{1}{\lambda}
    \log\left(\frac{1+\delta}{\delta}\right)\right),
    \end{align}
    \begin{align}\label{Tpro_upper_corollary2}
     T_{\text{pro}}^{\text{up}} = \frac{1}{p_{\text{pro}}^2}\left(1+\frac{2}{\lambda}\log\left(\frac{1+\delta+\sqrt{1+\delta}}{\delta}\right) \right).
\end{align}
\end{corollary}
{\em Proof:} 
The proof is provided in Appendix C. 
\hfill $\Box$

 \subsection{\label{sec:storage} Storage Analysis}
In this section, we provide storage requirements of the codes that we explained
in Section~\ref{sec:existingCodes}. These codes have storage requirements both at the master and worker devices. In particular, we assume that each entry of matrices $A$ and $B$ requires a fixed amount of memory. Our analysis, which is provided next, quantifies how many of these entries are needed to be stored at the master and worker devices.  

\subsubsection{Storage at Master} In all codes, we first store matrices $A$ and $B$, where each matrix contains $KL$ components. Also, we store the final result, $A^TB \in \mathbb{R}^{K \times K}$, which contains $K^2$ components. Therefore, $2KL + K^2$ entries should be stored at the master device for each code. 

In repetition codes, we store the first $k_{\text{rep}}$ results obtained from $k_{\text{rep}}$ workers, where $k_{\text{rep}} = N-\frac{N}{p_{\text{rep}}}+1$. Each result contains $\frac{K^2}{p_{\text{rep}}}$ components. 
\begin{align}
    S_{\text{rep}}^{m} = k_{\text{rep}}\frac{K^2}{p_{\text{rep}}}+2KL+K^2.
\end{align}

In MDS codes, we store $N-k_{\text{mds}}$ coded sub-matrices. Each coded matrix contains $\frac{KL}{p_{\text{mds}}}$ components. Then, we need to store the first $k_{\text{mds}}$ results obtained from $k_{\text{mds}}$ workers. Each result contains $\frac{K^2}{p_{\text{mds}}}$ components. 
\begin{align}
    S_{\text{mds}}^{m} = (N-k_{\text{mds}})\frac{KL}{p_{\text{mds}}}+k_{\text{mds}}\frac{K^2}{p_{\text{mds}}}+2KL+K^2.
\end{align}

In polynomial codes, we store $2N$ coded sub-matrices. Each coded matrix contains $\frac{KL}{p_{\text{poly}}}$ components. Then, we need to store the first $k_{\text{poly}}$ results obtained from $k_{\text{poly}}$ workers. Each such result contains $\frac{K^2}{p_{\text{poly}}^2}$ components. 
\begin{align}
    S_{\text{poly}}^{m} = 2N\frac{KL}{p_{\text{poly}}}+k_{\text{poly}}\frac{K^2}{p_{\text{poly}}^2}+2KL+K^2.
\end{align}

In MatDot, we store $2N$ coded sub-matrices. Each coded matrix contains $\frac{KL}{p_{\text{matdot}}}$ components. Then, we need to store the first $k_{\text{matdot}}$ results collected from $k_{\text{matdot}}$ workers. Each result contains $K^2$ components. 
\begin{align}
    S_{\text{matdot}}^{m} = 2N\frac{KL}{p_{\text{matdot}}}+k_{\text{matdot}}K^2+2KL+K^2.
\end{align}

In product codes, we store $2(N-k_{\text{pro}})$ coded sub-matrices. Each coded matrix contains $\frac{KL}{p_{\text{pro}}}$ components. Then, we need to store the first $k_{\text{pro}}$ results collected from $k_{\text{pro}}$ workers, where $k_{\text{pro}} = 2(p_{\text{pro}}-1)\sqrt{N}-{(p_{\text{pro}}-1)}^2+1$ \cite{yu2017polynomial}. Each result contains $\frac{K^2}{p_{\text{pro}}^2}$ components. 
\begin{equation}
    S_{\text{pro}}^{m} = 2(N-k_{\text{pro}})\frac{KL}{p_{\text{pro}}}+k_{\text{pro}}\frac{K^2}{p_{\text{pro}}^2}+2KL+K^2.
\end{equation}

\subsubsection{Storage at Workers}
In repetition codes, each worker receives one matrix of size $\frac{KL}{p_{\text{rep}}}$ and one matrix of size $KL$, as we decompose only one of the matrices to $p_{\text{rep}}$ parts. So, the size of resulting matrix is $\frac{K^2}{p_{\text{rep}}}$. Similarly, in MDS codes we decompose only one of the matrices to $p_{\text{mds}}$ parts. Each worker receives one matrix with size of $\frac{KL}{p_{\text{mds}}}$ and one matrix with size of $KL$. Thus, the size of resulting matrix is $\frac{K^2}{p_{\text{mds}}}$. Thus, the storage requirement of repetition and MDS codes is expressed as 
\begin{align} \label{eq:storage_rep_mds}
    S_{x}^{w} = \frac {KL}{p_{x}}+KL+\frac{K^2}{p_{x}}, \forall x \in \{\text{rep},\text{mds}\}. 
\end{align}  

In polynomial and product codes, each worker receives two matrices of size $\frac{KL}{p_{x}}$, $x \in \{\text{poly}, \text{pro}\}$. The size of the matrix after multiplication is  $\frac{K^2}{p_{x}^2}$. Therefore, the storage requirement of polynomial and product codes at each worker is 
\begin{align} \label{eq:storage_poly_pro}
    S_{x}^{w} = \frac {2KL}{p_{x}}+\frac{K^2}{p_{x}^2}, \forall x \in \{\text{poly}, \text{pro}\}.
\end{align}

On the other hand, the size of the matrix after computation is $K^2$ in MatDot as matrix partitioning is done differently (\ie row-wise, which means that $A_i \in \mathbb{R}^{\frac{L}{p_{\text{matdot}}} \times K}$ and $B_i \in \mathbb{R}^{\frac{L}{p_{\text{matdot}}} \times K}$ after partitioning and finally, $A^{T}_iB_i \in \mathbb{R}^{K \times K} $) as compared to polynomial and product codes (partitions column-wise, which means that we have $A_i \in \mathbb{R}^{L \times \frac{K}{p_{x}}}$ and $B_i \in \mathbb{R}^{L \times \frac{K}{p_{x}}}$ after partitioning, 
and the final result is $A^{T}_iB_i \in \mathbb{R}^{\frac{K}{p_{x}} \times \frac{K}{p_{x}}}, \forall x \in \{\text{poly}, \text{pro}\} $). Therefore, the storage requirement of MatDot is expressed as 
\begin{align} \label{eq:storage_matdot}
    S_{\text{matdot}}^{w} = \frac{2KL}{p_{\text{matdot}}}+K^2.
\end{align}

\subsection{\label{sec:designACMM} Design of \ACMM Algorithm} In this section, we present our  \ACMM algorithm. We consider an iterative process such as gradient descent, where matrix multiplications are required at each iteration. Our goal is to determine the best matrix multiplication code and the optimum number of matrix partitions by taking into account the task completion delay, \ie computing time, storage requirements and decoding probability of each code. In particular, \ACMM solves an optimization problem at each iteration, and determines which code is the best as well as the optimum number of partitions for that code. For example, MDS codes may be good at iteration $i$, while polynomial codes may suit better in later iterations. The optimization problem is formulated as
\begin{align} \label{eq:opt}
    \min_{\pi, p_{\pi}} \text{\; \;} & T_{\pi} \nonumber \\
 \text{subject to \; \;} & S_{\pi}^{z} \leq S_{\text{thr}}^{z}, z \in \{m,w\}, \nonumber \\
 & \rho_{\pi} \geq \rho_{\text{thr}}, \nonumber \\
 & k_{\pi} \leq N, \nonumber \\
 & p_{\pi} \geq 2, \nonumber \\
 & \pi \in \{\text{rep}, \text{mds}, \text{poly}, \text{matdot}, \text{pro}\}.
\end{align} The objective function selects the best code $\pi$ from the set $\{\text{rep}, \text{mds}, \text{poly}, \text{matdot}, \text{pro}\}$ as well as the optimum number of partitions $p_{\pi}$. The first constraint is the storage constraint, which limits the storage usage at master and worker devices with thresholds $S_{\text{thr}}^{m}$ and $S_{\text{thr}}^{w}$. The second constraint is the successful decoding constraint, where successful decoding probability   $ \rho_{\pi}$ should be larger than the threshold  $\rho_{\text{thr}}$. The successful decoding probability is defined as the probability that the master receives all the required results from workers. If we assume that the failure probability of each worker is $(1-\phi)$ and independent, the total number of workers is $N$ and the number of sufficient results is $k_\pi$, one may formulate the probability of success for each coding method as a binomial probability $\rho_\pi = \sum_{i=k_\pi}^{N} \binom{N}{i}  (\phi)^i(1-\phi)^{N-i}$. The third constraint makes sure that the recovery threshold $k_\pi$ is less than the number of workers. The fourth constraint makes the number of partitions larger than or equal to 2, otherwise there is no matrix partitioning, which is a degenerate case. 

 \section{\label{sec:simulation} Performance Analysis of \ACMM}

In this section, we evaluate the performance of our algorithm; \ACMM via simulations. 
We consider a master/worker setup, where per sub-matrix computing delay $\lambda$ is an i.i.d. random variable following a shifted 
exponential distribution. We compare \ACMM with the baselines; repetition, MDS, polynomial, MatDot, and product codes. 

\begin{figure}
		\centering
		\includegraphics[width=8cm]{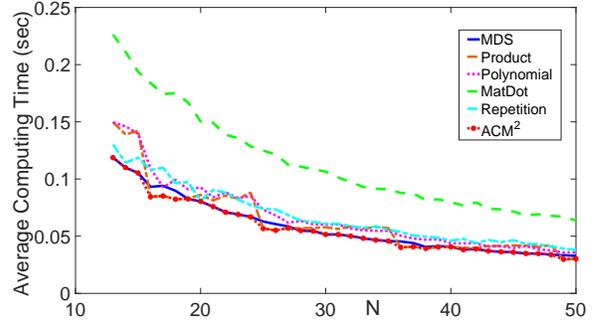}
		\label{label2}
	\caption{Task completion delay. \textbf{There are no storage or successful decoding probability constraints}. 
}
\label{fig:noCons_mu_greater_1}
\vspace{-5pt}
\end{figure}

Fig.~\ref{fig:noCons_mu_greater_1} shows the average computing time versus number of workers $N$ when there is no storage or successful decoding probability constraint in (\ref{eq:opt}). In this setup, $K=2000$, $L=5000$, $\phi = 0.95$, $\lambda$ is randomly and uniformly selected from  $\{2,3,4,5,6,7,8,9,10\}$. 
In this setup, workers are fast (since $\lambda > 1$), so all coding algorithms prefer splitting matrices to more partitions. This causes low computing load at workers, but the master needs to wait for more results, \ie recovery threshold $k$ is large, to be able to decode computed sub-matrices. In this setup, the optimum number of partitions of repetition codes is equal to $N$ ($p^*_{\text{rep}} = N$), while for MDS codes it is close to $N$ ($0.9<\frac{p^*_{\text{mds}}}{N}<1$). This means that repetition codes are the same as no coding and MDS gets closer to no coding  as $\lambda$ increases \cite{SpeedUp-conference,SpeedUp-journal}. When $N$ is small,  
no coding is the best, so \ACMM chooses MDS codes (which is close to repetition codes) as they behave as no coding. When $N$ increases, MDS, polynomial, and product codes perform better than repetition codes. 
In this setup, product codes operate in the first regime, because workers are fast. Thus, the optimum number of partitions is large and close to $N$, so, $N = p_{\text{pro}}^2 + \tau p_{\text{pro}}$. 
Product codes perform better in this regime \cite{lee2017high}.  
When $\sqrt N$ is integer, product codes are the best as they can use all existing workers and choose the number of partitions as large as possible to decrease the computation load of each worker. However, when $\sqrt N$ is not integer, product codes may waste resources of some workers (as they only use $\lfloor \sqrt N \rfloor$ workers), so MDS and polynomial codes perform better. Computation time of MDS is less than or equal to polynomial, because computation load of polynomial is higher than MDS codes. 
The optimum number of partitions for each code increases with increasing $N$, which decreases the computation load at each worker. Since all workers are fast ($\lambda > 1$), all codes choose as large partitions as possible. Thus, they perform close to each other. 
MatDot performs worse than the other codes, because of its different way of partitioning; \ie row-wise versus column-wise. Thus, MatDot introduces almost $2$ times more computation load. 
As seen, \ACMM exploits the best code among all codes, so it performs the best. 


\begin{figure}
		\centering
		\includegraphics[width=8cm]{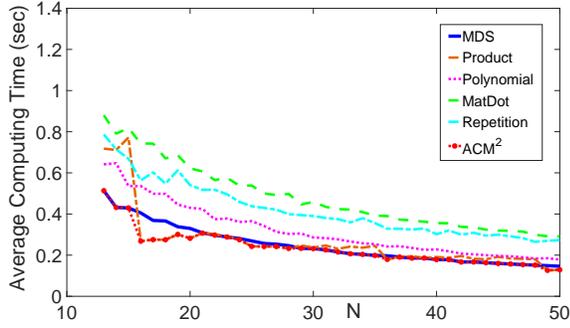}
	\caption{Task completion delay. \textbf{Storage is constrained}. }
\label{fig:SCons_mu_greater_1}
\end{figure}

Fig.~\ref{fig:SCons_mu_greater_1} demonstrates average computing time versus number of workers when there exists storage constraint in (\ref{eq:opt}). In this setup, $K=2000$, $L=5000$, $\phi = 0.95$, $\lambda$ is selected randomly and uniformly from $\lambda \in \{\frac{1}{10}, \frac{1}{9}, \frac{1}{8}, \frac{1}{7}, \frac{1}{6}, \frac{1}{5}, \frac{1}{4}, \frac{1}{3}, \frac{1}{2}\}$, and the storage constraint is set to $S_{\text{thr}}^{w}=15$M entries. 
%
%
In this scenario, as the workers are slow, \ie $\lambda < 1$, all codes prefer to choose small number of partitions. There is a trade-off between the number of partitions and storage requirement. It means that the storage requirement reduces with increasing number of partitions as smaller matrices are multiplied by each worker, so less storage is needed. Since there is a storage constraint, all codes prefer to increase the number of partitions. \ACMM exploits this trade-off and selects the best code and optimum number of partitions. 



\begin{figure}
		\centering
		\includegraphics[width=8cm]{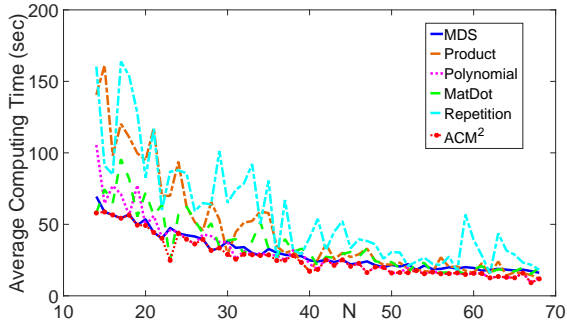}
	\caption{Task completion delay. \textbf{Storage and success probability are constrained}. }
\label{fig:PSandSCons_mu_less_1}
\vspace{-15pt}
\end{figure}

Fig.~\ref{fig:PSandSCons_mu_less_1} illustrates average computing time versus number of workers when there exists both storage and success probability constraints in (\ref{eq:opt}). In this setup, $K=2000$, $L=5000$, $\phi = 0.9$, $\lambda$ is selected randomly and uniformly from $\lambda \in \{\frac{1}{2000},\frac{1}{1000}, \frac{1}{900}, \frac{1}{800}, \frac{1}{700}, \frac{1}{600}, \frac{1}{500}\}$, the storage constraint is set to $S_{\text{thr}}^{w}=10$M entries and the success probability constraint is set to $\rho_{\text{thr}} = 0.98$. In this scenario, our proposed algorithm selects any of the MDS, product, polynomial, MatDot and repetition codes at least one time during these iterations. In general in this setup, MatDot and polynomial codes perform better as compared with Fig.~\ref{fig:noCons_mu_greater_1} and Fig.~\ref{fig:SCons_mu_greater_1}. Polynomial codes work better due to the tighter storage constraint and MatDot codes perform better because of the existence of success probability constraint, which has an inverse relation with the recovery threshold.

\section{\label{sec:conclusion} Conclusion}
In this paper, we focused on characterizing the cost-benefit trade-offs of coded computation for practical edge computing systems, and develop an adaptive coded computation framework. In particular, we studied matrix multiplication as a computationally intensive task, and developed an adaptive coding for matrix multiplication (\ACMM) algorithm by taking into account the heterogeneous and time varying nature of edge devices. \ACMM dynamically selects the best coding policy by taking into account the computing time, storage requirements as well as successful decoding probability. 

 
 \bibliographystyle{IEEEtran}

 \bibliography{ACMM}
 
 \section{\label{sec:appendixa} Appendix A: Construction of Table~\ref{table:codes_comparison}} 
In this appendix, we provide detailed calculations as well as some definitions related to Table~\ref{table:codes_comparison}. 

\subsection{Recovery Threshold $(k)$}

Recovery threshold is defined as the minimum number of results that the master needs to receive to be able to compute the final result. In product codes the master can finish matrix multiplication calculation if it receives any $k_{\text{pro}} = 2(p_{\text{pro}} - 1)\sqrt{N} - (p_{\text{pro}} - 1)^2 +1$ results from workers \cite{yu2017polynomial}. 
In the example of Table~\ref{table:codes_comparison}, by plugging in $N = 9$ and $p_{\text{pro}} = 2$, the recovery threshold for product code becomes $k_{\text{pro}} = 6$. In polynomial codes, $k_{\text{poly}} = p^2_{\text{poly}} = 4$. Indeed, there are four unknowns in $\alpha(n)\beta(n) = A_0B_0 + A_1B_0n + A_0B_1n^2 + A_1B_1n^3$, so the master needs to receive at least four results from workers to be able to calculate $C=A^TB$. 
In MatDot codes, $k_{\text{matdot}} = 2p_{\text{matdot}}-1 = 3$. Since there exist three unknowns in $\alpha(n)\beta(n) = (A_0 + A_1n)(B_0n + B_1) = A_0B_1 + (A_0B_0 + A_1B_1)n + A_1B_0 n^2$, the master needs to receive at least three results from workers to calculate $C=A^TB$. In MDS codes, $k_{\text{mds}}=p_{\text{mds}}=2$ \cite{SpeedUp-journal}, and in repetition codes $k_{\text{rep}} = N-\frac{N}{p_{\text{rep}}}+1$, for $p_{\text{rep}}|N$, since the master can decode the final result when it receives $A_i^TB$, $\forall i=1, \ldots, p_{\text{rep}}$. Therefore, in the example of Table~\ref{table:codes_comparison}, by plugging in $p_{\text{rep}}=2$, the recovery threshold for repetition codes become $\lfloor \frac{N}{2}+1 \rfloor$.
%

\subsection{Computing Load per Worker ($\gamma$)}

We define the computing load as the total number of multiplications (\ie $a_{i,k}b_{k,j}$, $\forall i,k,j$) to compute $C=A^TB$. We note that the computing load could be considered as the static version of the computing time analysis. We assume that each  $a_{i,k}b_{k,j}$ multiplication takes a fixed amount of time, so the total computing time will be proportional to the the number of such multiplications . 
Next, we provide 
the detailed calculations for computing load per worker. 

In product codes, the matrix partitioning is column-wise, so we have $A_i \in \mathbb{R}^{K \times \frac{K}{2}}$ and $B_i \in \mathbb{R}^{K \times \frac{K}{2}}$, for $ i \in \{0,1\}$, after partitioning. Thus, $\frac{K}{2} \times K \times \frac{K}{2} = \frac{K^3}{4}$ multiplications are needed to compute $A^{T}_iB_i \in \mathbb{R}^{\frac{K}{2} \times \frac{K}{2}}$. In polynomial codes, the matrix partitioning is also column-wise, and we have $A_i \in \mathbb{R}^{K \times \frac{K}{2}}$ and $B_i \in \mathbb{R}^{K \times \frac{K}{2}}$, for $ i \in \{0,1\}$ after partitioning, so $\alpha(n) \in \mathbb{R}^{\frac{K}{2} \times K}$ and $\beta(n) \in \mathbb{R}^{K \times \frac{K}{2}}$. Similar to the product codes, $\frac{K}{2} \times K \times \frac{K}{2} = \frac{K^3}{4}$  multiplications are needed to compute $\alpha(n)\beta(n) \in \mathbb{R}^{\frac{K}{2} \times \frac{K}{2}}$. In MatDot codes, the matrix partitioning is row-wise. It means that we have $A_i \in \mathbb{R}^{\frac{K}{2} \times K}$ and $B_i \in \mathbb{R}^{\frac{K}{2} \times K}$, for $ i \in \{0,1\}$ after partitioning. Hence, $\alpha(n) \in \mathbb{R}^{K \times \frac{K}{2}}$ and $\beta(n) \in \mathbb{R}^{\frac{K}{2} \times K}$ and $K \times \frac{K}{2} \times K = \frac{K^3}{2}$ multiplications are needed to compute $\alpha(n)\beta(n) \in \mathbb{R}^{K \times K}$. In both MDS and repetition codes, one of the matrices, say $A$, is partitioned column-wise, so we have $A_i \in \mathbb{R}^{K \times \frac{K}{2}}$, for $ i \in \{0,1\}$, after partitioning and $B \in \mathbb{R}^{K \times K}$. Thus, $\frac{K}{2} \times K \times K = \frac{K^3}{2}$ multiplications are needed to compute $A^{T}_iB \in \mathbb{R}^{\frac{K}{2} \times K}$.


\subsection{Storage Load Per Worker $(\mu)$}

We can compute the storage load per worker of MDS, repetition, product, polynomial and MatDot codes according to (\ref{eq:storage_rep_mds}), (\ref{eq:storage_poly_pro}) and (\ref{eq:storage_matdot}). 
%
In the example of Table~\ref{table:codes_comparison}, assuming $K = L$ and $p_{\text{mds}}=p_{\text{rep}}=p_{\text{poly}} = p_{\text{pro}} = p_{\text{matdot}} = 2$, storage load per worker of polynomial and product codes becomes $K^2 + \frac{K^2}{4}$, while it is $2K^2$ for MDS, repetition and MatDot codes.


\subsection{Probability of successful computation $(\rho)$}
As we discussed in Section~\ref{sec:designACMM}, the successful decoding probability can be defined as $\rho_\pi = \sum_{i=k_\pi}^{N} \binom{N}{i}  (\phi)^i(1-\phi)^{N-i}$, where $(1-\phi)$ is the failure probability of each worker.
By plugging in $(1-\phi) = \frac{1}{3}$, $k_{\text{pro}} = 6$, $k_{\text{poly}} = 4$, $k_{\text{matdot}} = 3$, $k_{\text{mds}} = 2$ and $k_{\text{rep}} = \lfloor \frac{N}{2}+1 \rfloor$ we can compute the successful decoding probability for different $N \in \{6,7,8,9\}$ for each coding method in Table~\ref{table:codes_comparison}.

\section{\label{sec:appendixb} Appendix B: Proof of Corollary~\ref{polymatdottime}}

 
 \subsection{Expected Value of the $k^{\text{th}}$ Order Statistic of Exponential Distribution}
 
 \begin{definition}
 \label{osdefinition}
Assume $X_1,X_2,\dots,X_n$ are any $n$ real valued random variables. If $X_{(1)} \leq X_{(2)} \leq\dots\leq X_{(n)}$ represent the ordered values of $X_1,X_2,\dots,X_n$, then, $X_{(1)},X_{(2)},\dots,X_{(n)}$ will be called the order statistics of $X_1,X_2,\dots,X_n$ \cite{DasGupta2011}. 
\end{definition}

\textit{Reyni’s Representation}: The order statistics of an exponential distribution can be shown as linear combinations of independent exponential random variables with a special sequence of coefficients
 \begin{equation}
     X_{(k)} \overset{L}{=} \sum_{i = 1}^{k}\frac{X_i}{n-i+1}, 
 \end{equation}
 for $k = 1,\dots,n$. Where $\overset{L}{=} $ means equal in distribution and $X_1,\dots,X_n$ are independent exponential random variables with mean $\frac{1}{\lambda}$ \cite{DasGupta2011}. 
 
The expected value of the $k^{\text{th}}$ order statistic of exponential distribution with mean $\frac{1}{\lambda}$, is expressed as the following
\begin{align}
E[X_{(k)}] = \frac{1}{\lambda}(H_n - H_{(n-k)}),
\end{align}
where $H_n \triangleq \sum_{i = 1}^{n}\frac{1}{i}$.


  
  \subsection{Expected Value of the $k^{\text{th}}$ Order Statistic of Shifted-Scaled Exponential Distribution}

 The function $f(X) = a + bX$ is a \emph{monotone} function as its first derivative does not change sign. When a monotone function is applied to an
 ordered set, it preserves the given order \cite{TheConciseOxfordDictionaryofMathematics}. 
 Now, let us define $Y_i = a + bX_i$ for $i  = 1,\dots,n$ where $X_i$ are independent exponential random variables with mean $\frac{1}{\lambda}$. By monotonicity, we have $Y_{(k)} = a + bX_{(k)}$ and by considering \textit{Reyni’s} representation for the $k^{\text{th}}$ order statistic of exponential random variables, we have 
  \begin{equation}
     Y_{(k)} \overset{L}{=} a+\sum_{i = 1}^{k}\frac{bX_i}{n-i+1}. 
 \end{equation}
 
The expected value of the $k^{\text{th}}$ order statistic of shifted-scaled exponential distribution with mean $a + \frac{b}{\lambda}$ is expressed as the following
\begin{align}\label{Cor4equation}
E[Y_{(k)}] = a + \frac{b}{\lambda}(H_n - H_{(n-k)}).
\end{align}


  \subsection{Proof of Corollary~\ref{polymatdottime}}
  In this paper, we assume shifted exponential distribution for computation time of the original task and shifted-scaled exponential distribution for computation time of each sub-task. Thus, if we assume $X$ as an exponential random variable with mean $\frac{1}{\lambda}$, the computation time of the original task can be expressed as $t = X + 1$ and if we assume $X_i$, $i = 1,\dots,N$ as $N$ independent exponential random variables with mean $\frac{1}{\lambda}$, the computation time of each sub-task can be expressed as $t_i = \frac{X_i + 1}{\alpha}, \alpha \in \{p_{\text{rep}}, p_{\text{mds}}, p^2_{\text{poly}}, p_{\text{matdot}}, p^2_{\text{pro}}\}$. Hence, if we replace both $a$ and $b$ with $\frac{1}{\alpha}$ in (\ref{Cor4equation}), the expected value of the computation time of each worker for polynomial and MatDot codes can be shown as $\frac{1}{\alpha}+\frac{1}{\alpha\lambda}(H_N-H_{(N-k_i)})$ where, $ k_i \in \{k_{\text{poly}}, k_{\text{matdot}}\}$ and $\alpha \in \{p^2_{\text{poly}}, p_{\text{matdot}}\}$. Moreover, the $N^{\text{th}}$ harmonic number, $H_N$ can be approximated by natural logarithm function, i.e.,
   $H_N\approx log(N)$ and $H_{(N-k_i)}\approx log(N-k_i)$. Therefore, the expected value of the computation time of each worker for aforementioned coding methods becomes
  \begin{equation}
      T_i \approx \frac{1}{\alpha}(1+\frac{1}{\lambda}log(\frac{N}{N-k_i})),
  \end{equation}
  where $i \in \{\text{poly}, \text{matdot}\}$, $ k_i \in \{k_{\text{poly}}, k_{\text{matdot}}\}$ and $\alpha \in \{p^2_{\text{poly}}, p_{\text{matdot}}\}$.
  \hfill $\Box$

\section{\label{sec:appendixc} Appendix C: Proof of Corollary~\ref{producttime}}
 \subsection{Asymptotic Computation Time of Product Codes }\label{partAappendixC}
 Although we can directly use order statistics to compute the  computation time of repetition, MDS, polynomial and MatDot codes, it is challenging for product codes, because the decodability condition of product codes depends on which specific tasks are completed.  However, with the help of the idea of edge-removal process in a bipartite graph \cite{4313069} and order statistics, one can find an asymptotic computation time for product codes in two different regimes. 
Assuming exponential distribution as a delay model with mean $\frac{1}{\lambda}$ for each worker, the average computing time $T^{'}_{\text{pro}}$ for products codes in the first regime is as the following \cite{lee2017high, 4313069, PITTEL1996111}

\begin{align}\label{T_pro first regime}
    T^{'}_{\text{pro}} \approx \frac{1}{\lambda}\log\left(\frac{p_{\text{pro}}+\frac{\tau}{2}}{c_{\tau/2+1}}\right),
\end{align} where $c_{\tau/2+1} \approx (1+\tau/2)+\sqrt{(1+\tau/2)\log(1+\tau/2)}$.\\

 Also, the average computation time of $(\sqrt{1+\delta}p_{\text{pro}}, p_{\text{pro}})^2$ product codes with $(1 + \delta)p^2_{\text{pro}}$ workers (assuming exponential distribution with mean $\frac{1}{\lambda}$ for the computation time of each worker), for a fixed constant $\delta$, as $p_{\text{pro}}$ grows to infinity, is lower bounded as follows \cite{lee2017high}
\begin{align}\label{T_pro lower bound}
    T'_{\text{pro}} \geq \frac{1}{\lambda} \log\left(\frac{1+\delta}{\delta}\right).
\end{align}
On the other hand, since we consider the worst case scenario to compute the recovery threshold of product codes \cite{yu2017polynomial}, the upper bound on the computational time of the product codes in the second regime is the $(k_{\text{pro}})^{\text{th}}$ order statistics of $(1 + \delta)p^2_{\text{pro}}$ computational times, where $k_{\text{pro}} = N - (\sqrt{N} - p_{\text{pro}}+1)^2 +1$.
\begin{align} \label{T_pro upper bound}
    T'_{\text{pro}} \leq \frac{1}{\lambda} \log\left(\frac{N}{N - k_{\text{pro}}}\right)\\
     = \frac{1}{\lambda} \log\left(\frac{N}{(\sqrt{N} - p_{\text{pro}}+1)^2 -1}\right)\\
     = \frac{1}{\lambda}\left( \log\left(\frac{\sqrt{N}}{\sqrt{N} - p_{\text{pro}}}\right) + \log\left(\frac{\sqrt{N}}{\sqrt{N} - p_{\text{pro}}+2 }\right) \right)\\
     \leq \frac{1}{\lambda}\left( \log\left(\frac{\sqrt{N}}{\sqrt{N} - p_{\text{pro}}}\right) + \log\left(\frac{\sqrt{N}}{\sqrt{N} - p_{\text{pro}} }\right) \right) \\
     = \frac{2}{\lambda} \log\left(\frac{\sqrt{N}}{\sqrt{N} - p_{\text{pro}}}\right) \\
=  \frac{2}{\lambda}\log\left(\frac{(\sqrt{1+\delta})p_{\text{pro}}}{(\sqrt{1+\delta})p_{\text{pro}}- p_{\text{pro}}}\right)\\
= \frac{2}{\lambda} \log\left(\frac{1+\delta+\sqrt{1+\delta}}{\delta}\right).
\end{align}
Therefore, based on the above explanations, one can define $T'^{\text{up}}_{\text{pro}} := \frac{2}{\lambda} \log\left(\frac{1+\delta+\sqrt{1+\delta}}{\delta}\right)$ and  $T'^{\text{low}}_{\text{pro}} := \frac{1}{\lambda} \log\left(\frac{1+\delta}{\delta}\right)$ as upper bound and lower bound of computational time of product codes, with exponential distribution as delay model of each worker in the second regime. 


\subsection{Proof of Corollary~\ref{producttime}}
One can find asymptotic computation time of product codes assuming exponential distribution as a delay model for each worker in the previous section of this Appendix. On the other hand, we assume shifted-scaled exponential distribution for computation time of each worker in this paper. Therefore, according to equation~(\ref{Cor4equation}), we can compute (\ref{Tpro_corollary2}), (\ref{Tpro_lower_corollary2}) and (\ref{Tpro_upper_corollary2}) using (\ref{T_pro first regime}), (\ref{T_pro lower bound}) and (\ref{T_pro upper bound}), respectively by replacing both shift and scale parameters, $a$ and $b$, with $\frac{1}{p^2_{\text{pro}}}$ in equation~(\ref{Cor4equation}).
\hfill $\Box$
\IEEEtriggeratref{4}

\end{document}